# ENERGY PRODUCTION DEMONSTRATOR FOR MEGAWATT PROTON BEAMS[*][†]

Vitaly S. Pronskikh[1,#], Nikolai V. Mokhov[1], Igor Novitski[1], Sergey I. Tyutyunnikov[2]

[1]Fermi National Accelerator Laboratory, Batavia, IL 60510, USA
[2]Joint Institute for Nuclear Research, Dubna, 141980

## Abstract

A preliminary study of the Energy Production Demonstrator (EPD) concept - a solid heavy metal target irradiated by GeV-range intense proton beams and producing more energy than consuming - is carried out. Neutron production, fission, energy deposition, energy gain, testing volume and helium production are simulated with the MARS15 code for tungsten, thorium, and natural uranium targets in the proton energy range 0.5 to 120 GeV. This study shows that the proton energy range of 2 to 4 GeV is optimal for both a $^{nat}$U EPD and the tungsten-based testing station that would be the most suitable for proton accelerator facilities. Conservative estimates, not including breeding and fission of plutonium, based on the simulations suggest that the proton beam current of 1 mA will be sufficient to produce 1 GW of thermal output power with the $^{nat}$U EPD while supplying < 8% of that power to operate the accelerator. The thermal analysis shows that the concept considered has a problem due to a possible core meltdown; however, a number of approaches (a beam rastering, in first place) are suggested to mitigate the issue. The efficiency of the considered EPD as a Materials Test Station (MTS) is also evaluated in this study.

[*] Work supported by Fermi Research Alliance, LLC under contract No. DE-AC02-07CH11359 with the U.S. Department of Energy.

[†] Presented at the 12th Meeting on Shielding Aspects of Accelerators, Targets and Irradiation Facilities (SATIF-12), April 28-30, 2014, Fermi National Accelerator Laboratory, Batavia, Illinois, USA.

[#]vspron@fnal.gov

## Introduction

In this contribution we are considering a possibility for high-energy proton beams to be applied to build a demonstrator of energy production. The neutron production by proton accelerators was studied in USA in 1960s (for example, [1]). In 1970s a uranium target was considered by R.R. Wilson [2] for the Energy Doubler's 100 to 1000 GeV proton beams. A number of experimental [3-6] and simulation [7-9] studies have later been undertaken employing heavy metal and fissile targets. Most of the recent studies were devoted to the lead-bismuth liquid targets surrounded by blankets [10-11]. In this work, a solid target concept is considered.

## Neutron production and fission

We extend previous studies encompassing the 0.5 to 120 GeV proton energy range and simulating energy deposition in solid targets explicitly using the MARS15 code [12-13]. The model used in this work is shown in Figure 1. It is a 60 cm in radius and 110 cm long cylindrical target with a 10-cm diameter, and a 35-cm long beam entrance channel. Target and hole dimensions were chosen to keep the neutron leakage at the level of a few percent in the entire energy range (see Figure 2). The simulated proton beam is uniform and parallel, 10 cm in diameter.

**Figure 1. MARS15 model of the tungsten, thorium, and uranium targets.**

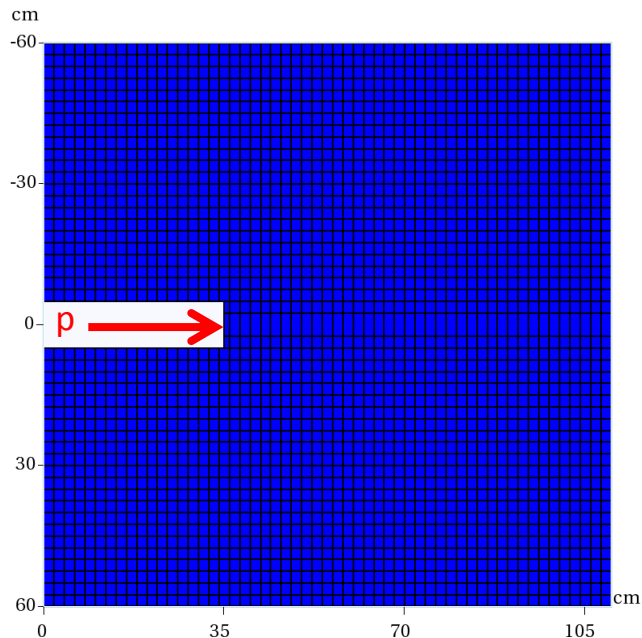

Three target materials - tungsten, natural uranium, and thorium - were studied. These are the good neutron producers as well as are quite abundant and representative of heavy fissile material. In the course of simulations, the CEM and LAQGSM event generators were invoked in MARS15 in order to determine outcomes of the hadron-nuclei interactions in the range (~few MeV – 8 GeV, for negative pions 0 – 8 GeV). Above 8 GeV for all interactions the MARS15 inclusive model is invoked. Figure 3 gives the number of neutrons generated in the target in all possible processes per incident proton and per GeV beam energy (i.e. energy cost of neutrons).

For all the three target materials these distributions show that the optimal energy for neutron production is between 2 and 4 GeV. The absolute number of neutrons per fission (Figure 4) is in agreement or slightly higher than in other studies for uranium [2], [8], and higher than for thorium [9]. Another difference is that many other studies report the optimal energy to be close to 1 GeV, while our simulations reveal the optimum at 2 - 4 GeV. This difference is most probably due to the fact that we use the latest version of LAQGSM generator for high-energy spallation which has recently been modified to incorporate a more detailed physics processes description [13] in the range 1 - 10 GeV based on a better adjustment of the model to all available differential data.

**Figure 2.  Leakage from the uranium target surface.**

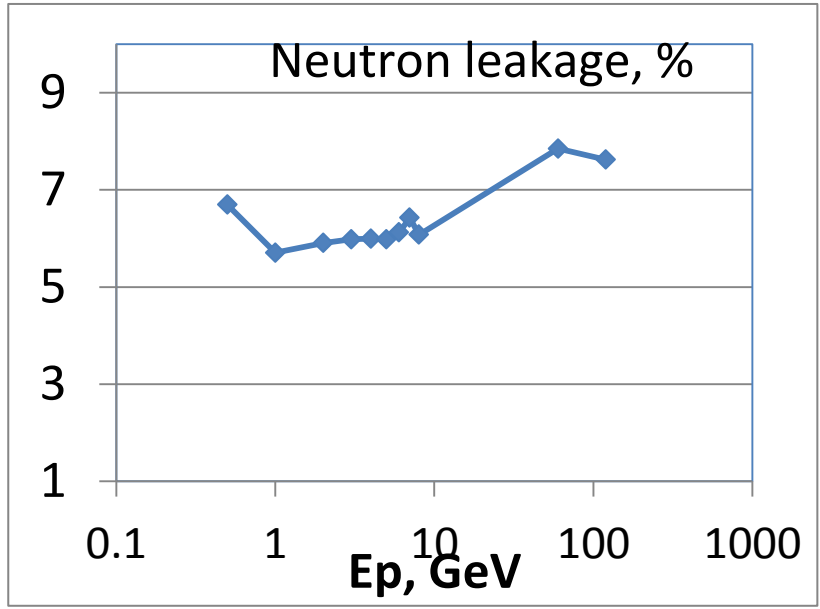

**Figure 3.  Number of neutrons released per one proton per GeV in the target.**

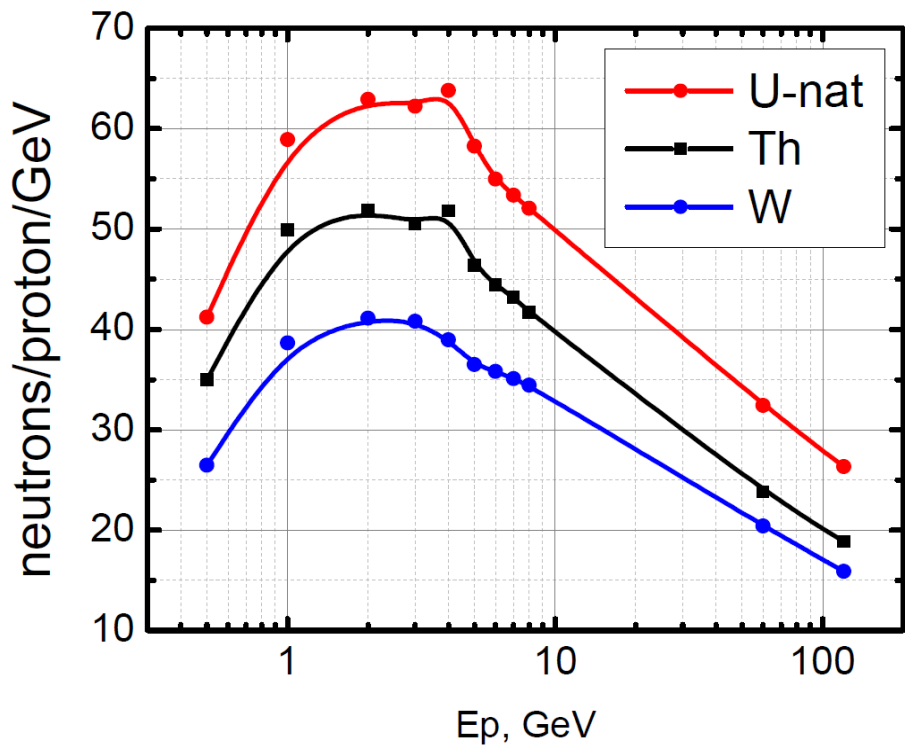

Most fissions in the uranium target (Figure 4) also occur in the above energy range, suggesting that a significant part of the neutrons are created in fission. In the case of the tungsten target the fission has peak at 1 GeV, but its contribution is smaller by three orders of magnitude. This explains the neutron surplus of ~20 neutrons in the uranium target as compared to the tungsten one.

**Figure 4. Number of fissions in the target per GeV proton energy.**

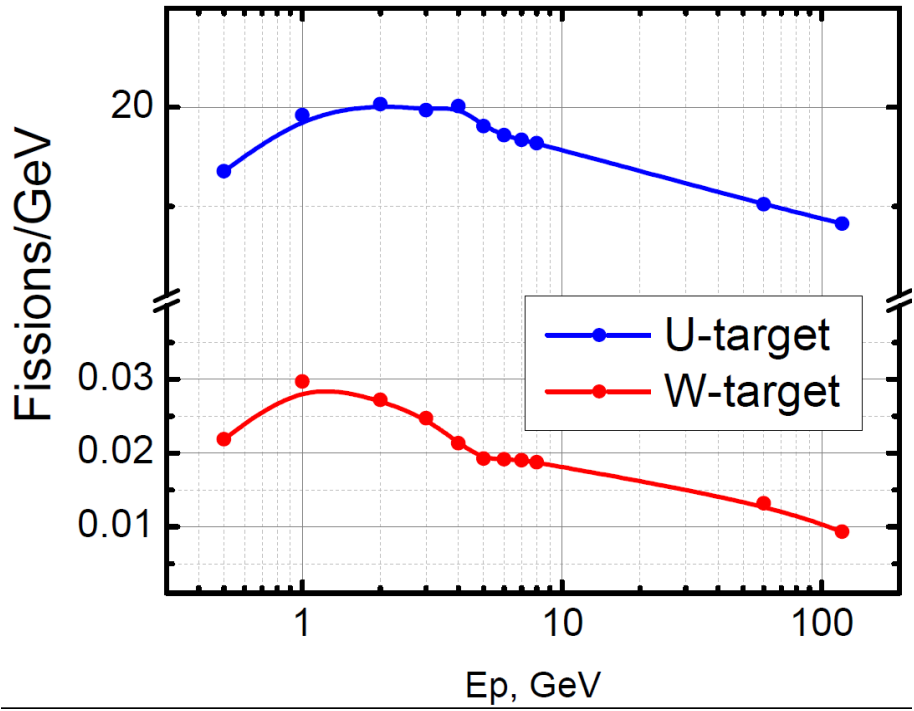

**Energy production**

Energy multiplication, which is the ratio of the energy deposited in the target to that of the primary beam impinging on it, is shown in Figure 5. It has the peak in the same 1 - 4 GeV range as above for natU because the factor is > 1 due to fission in the target. The fission cross section for tungsten is by orders of magnitude smaller than that for uranium-238. That is why the energy gain for tungsten is less than 1, and that material cannot serve efficiently for the energy production. The energy deposition in the thorium target is also much lower. Energy deposited in the target per neutron produced in it (Figure 6) is the quantity that shows the energy production efficiency as compared to neutron production efficiency at particular beam energy. This quantity has a minimum between 2 and 4 GeV for both natU and W; energy deposition in all processes in the target per one produced neutron is up to 6 times higher the natU target than in the W one. Note that for this quantity, the differences between its value at the optimal proton energy of 3 GeV and at minimal (0.5 GeV) and maximal (120 GeV) energies studied are about 15 %, making that difference not a very significant factor.

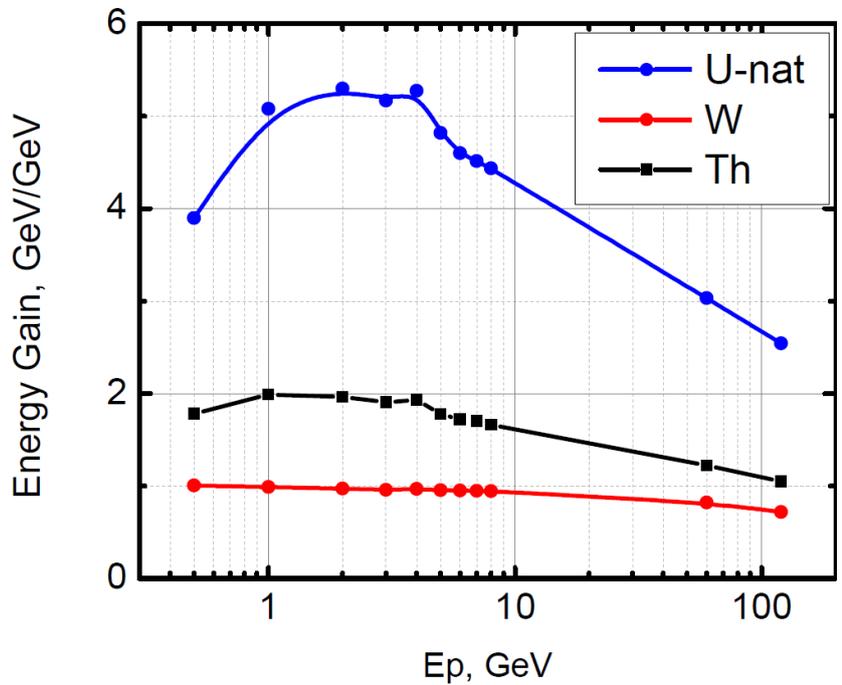

**Figure 5. Energy multiplication in the target**.

**Figure 6.  Energy released in the target per GeV proton energy.**

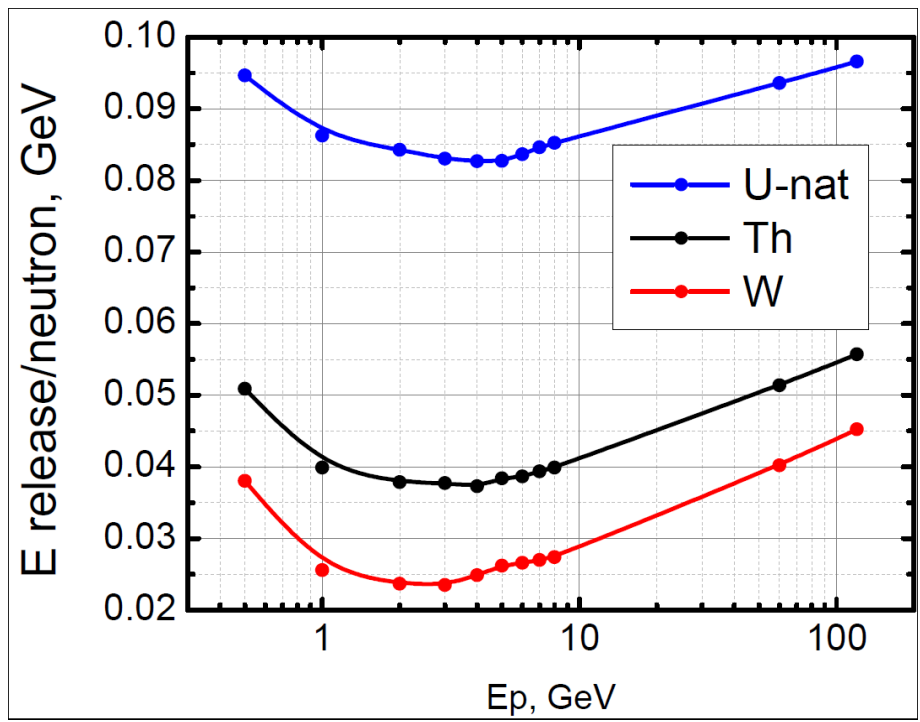

**Radiation damage and Materials Test Station**

The efficiency of the studied target as a Materials Test Station (MTS) is also evaluated here (see [11] for discussion of a liquid target concept proposed for that purpose). The aim of MTS is to maximize the DPA radiation damage as well as gas production in order to load the samples under study in the hottest location in the target. Figure 7 shows the target volume with DPA > 20 yr$^{-1}$ (at the beam intensity of $6.25 \cdot 10^{15}$ p/s), which is one of the reference numbers used to evaluate the MTS performance [11]. For both target materials such quantity has an optimum around 2 - 3 GeV.

**Figure 7. Testing volume in the target with DPA > 20 /yr per E$_p$**

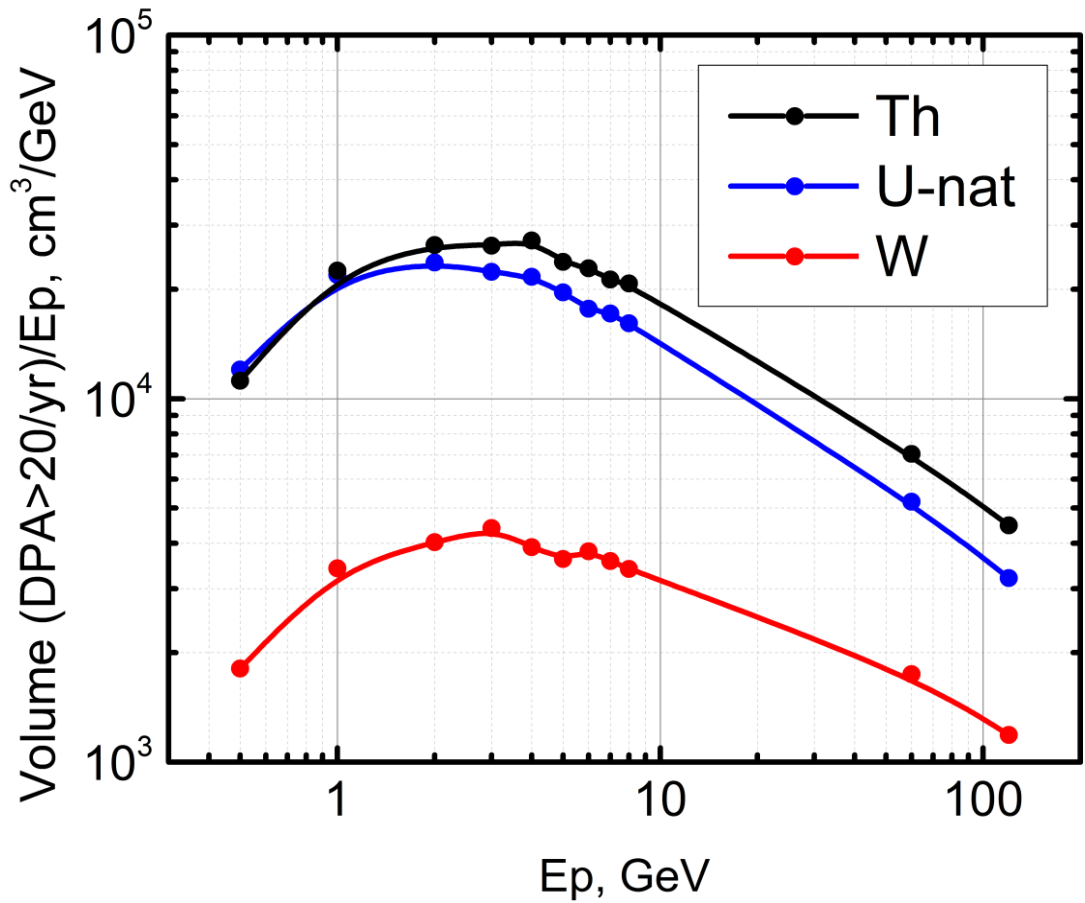

One of the key quantities to estimate radiation damage in a target or a reactor unit is the peak helium production expressed in units of appm/DPA per GeV beam energy (i.e. the energy cost of the gas production) (see Figure 8). It has a peak between 1 and 3 GeV; the helium production per unit energy grows faster than DPA up to 2 GeV, after which energy its growth becomes slightly slower. The Figure indicates that the gas/damage/E$_p$ ratio is highest (has the lowest cost) at lower energies between 0.5 and 4 GeV, while at 120 GeV it is two orders of magnitude less efficient. In absolute numbers, the helium production per DPA per E$_p$ is a factor of 7 less for the $^{nat}$U than for W target. For tungsten it is at the level of a typical fusion reactor and slightly less than for a spallation neutron source, like SINQ. For the $^{nat}$U target at 1 to 3 GeV it is at the level of a fission reactor on fast neutrons. The behaviour of the curve in Figure 8 above 8 GeV remains the same even if the LAQGSM model is used in that range instead of the inclusive one, for example, the appm/DPA/E$_p$ value at 120 GeV drops by ~20%.

**Figure 9. Peak helium production in the target, appm/DPA/E$_p$.**

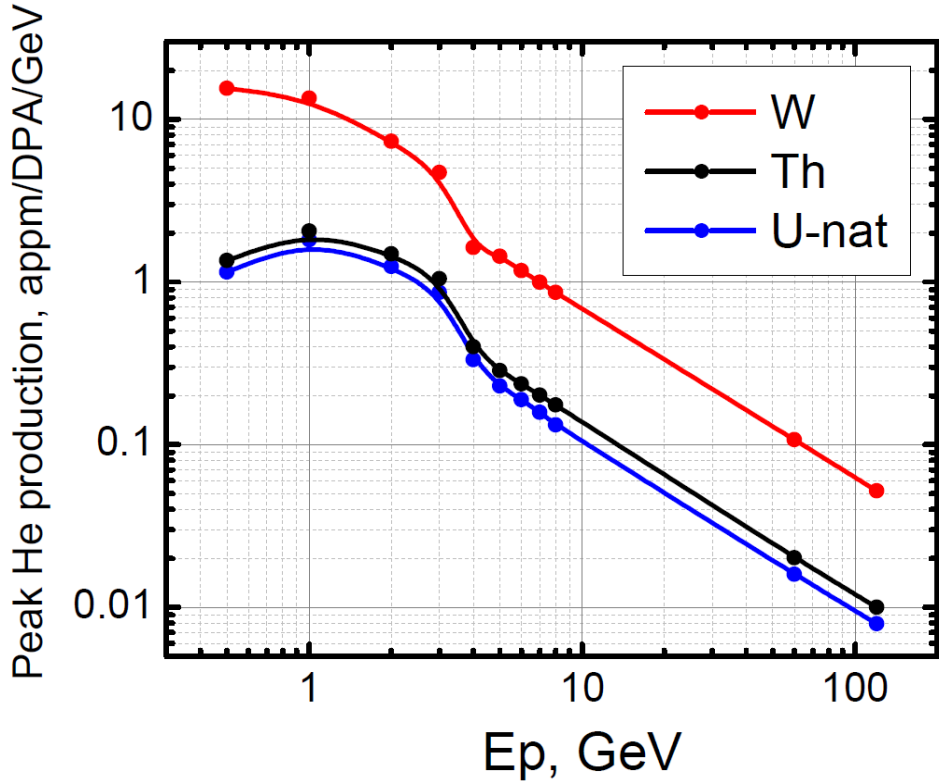

### Thermal output power

In order to use the target station in the energy production mode, so that the energy release in the target was higher than the energy used by accelerator, the following condition (1) should be satisfied:

$$\mathbf{P = P_{rel} - P_0 - P_{acc} \leq 0} \tag{1}$$

where P is the potential power produced by the station, $P_{rel}$ is the power released in the target, $P_0$ is the power needed to run accelerator in the idle mode, $P_{acc}$ is the power fed to RF system to accelerate protons.

It was estimated [2] that the proton intensity required at the Tevatron energies was obtained taking $P_0$ to be 20 MW, and $P_{acc} = b \cdot N \cdot E$, where $b = 2$. It is assumed that $P_{rel}$ is equal to 0.2 a$\cdot$N$\cdot$E, where 0.2 is the energy released per fission in GeV, and a $\approx$ 60 neutrons per fission, N is the proton beam intensity, and E is the proton energy. This latter is an estimate of the energy released in fission, assuming that neutron production is constant and each neutron is captured by uranium leading to the production of plutonium. A more detailed approach relevant to the accelerator-driven energy production calculations is described in [14]. In that approach, the thermal output power of an energy station can be described by the following equation (2):

$$\mathbf{P_0^{th} = I \cdot E_p \cdot G} \tag{2}$$

where I is the proton beam current (mA), $E_p$ is the proton beam energy (GeV), and G is the energy gain.

**Figure 10. Energy gain G.**

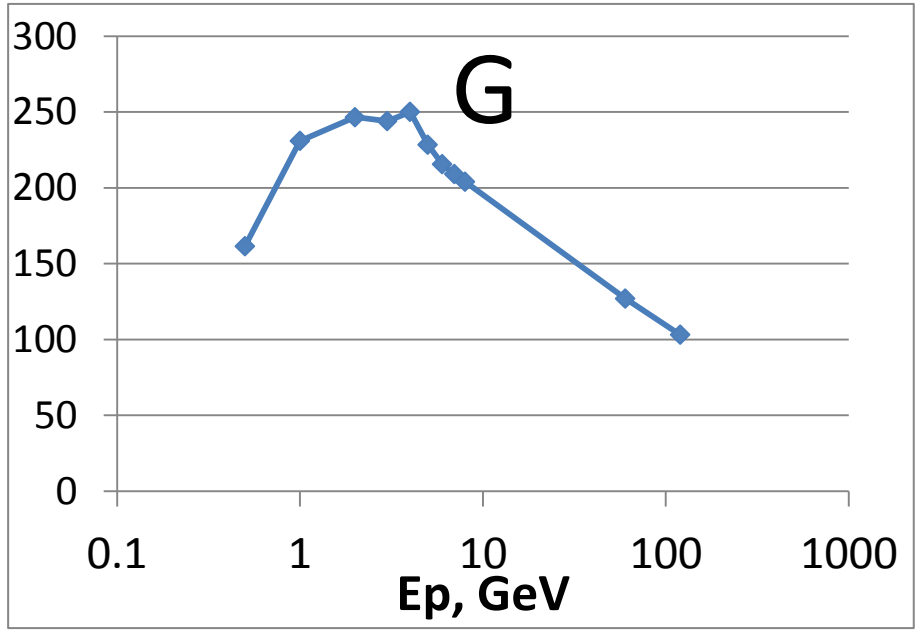

**Thermal output power**

In order to use the target station in the energy production mode, so that the energy release in the target was higher than the energy used by accelerator, the following condition (1) should be satisfied:

$$P = P_{rel} - P_0 - P_{acc} \leq 0 \qquad (1)$$

where P is the potential power produced by the station, $P_{rel}$ is the power released in the target, $P_0$ is the power needed to run accelerator in the idle mode, $P_{acc}$ is the power fed to RF system to accelerate protons.

It was estimated [2] that the proton intensity required at the Tevatron energies was obtained taking $P_0$ to be 20 MW, and $P_{acc}$ = b· N·E, where b = 2. It is assumed that $P_{rel}$ is equal to 0.2 a·N·E, where 0.2 is the energy released per fission in GeV, and a ≈ 60 neutrons per fission, N is the proton beam intensity, and E is the proton energy. This latter is an estimate of the energy released in fission, assuming that neutron production is constant and each neutron is captured by uranium leading to the production of plutonium. A more detailed approach relevant to the accelerator-driven energy production calculations is described in [14]. In that approach, the thermal output power of an energy station can be described by the following equation (2):

$$P_0^{th} = I \cdot E_p \cdot G \qquad (2)$$

where I is the proton beam current (mA), $E_p$ is the proton beam energy (GeV), and G is the energy gain.

The energy gain (see Figure 9) is the key quantity, a high value of which in a system allows a significant increase in the output power as compared to the proton beam power provided that the target is capable of an efficient neutron multiplication. The energy gain is described by equation (3) as follows:

$$G = \frac{\chi_s \cdot \varphi^* \cdot k_{eff} \cdot E_f}{\nu \cdot (1 - k_{eff})} \qquad (3)$$

where $\chi_s$ is the number of neutrons leaving the target and entering the blanket in target-blanket ADS; in our case the whole target (a "full absorption" target) serves its own blanket and both the neutron multiplication and energy production take place in its entire media and that is why in our case $\chi_s$ was taken to be equal to the number of neutrons produced per proton in the target (see Figure 3). The other quantities in the equation above were assumed to have the following values : $k_{eff} = 0.98$ (a typical number assumed for ADS), $\varphi^*=1$, neutron importance (can be larger than 1 if other neutron sources than fission exist in the system; in this case a conservative assumption is made), the number of fissions per neutron $\nu$ was taken to be 2.5, and $E_f$=0.2 GeV is the energy released per one fission.

Figure 10 gives the proton beam current required for a natural uranium target described in this work to produce 1 GW of thermal output power (Equation 1). It indicates that in the optimal energy range (assuming 4 GeV beam energy) one needs 1 mA of proton beam power to produce 1 GW output power; it requires more than 10 mA below 1 GeV, and it is at the level of 80μA for a 120-GeV beam.

Another important quantity is the fraction of the thermal output power required to support the accelerator operation. It is defined [14] by the equation (4):

$$f = \frac{1}{G \cdot \varepsilon \cdot \eta} \qquad (4)$$

where $\varepsilon$ is the electric to beam power conversion efficiency, 0.4, and $\eta$ is the thermal to electric power conversion efficiency, 0.45. Figure 11 shows the f dependence on the beam energy. The optimal beam energy (defined by the energy gain G) is also between 2 and 4 GeV.

**Figure 11. Beam current for 1 GW output power.**

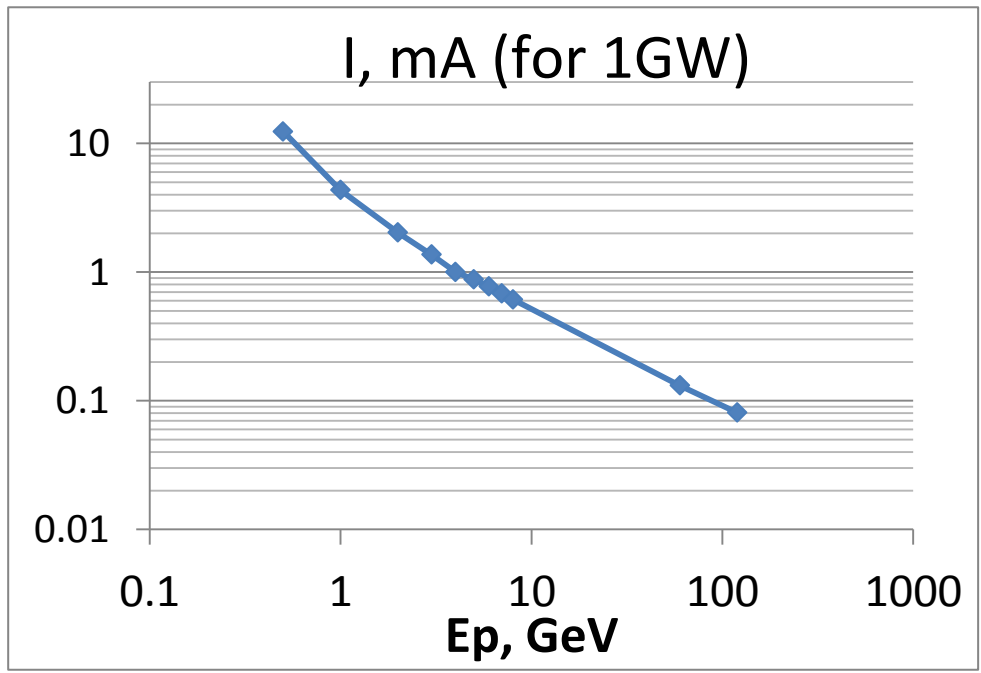

**Figure 12.  Fraction of the output power required to operate the accelerator.**

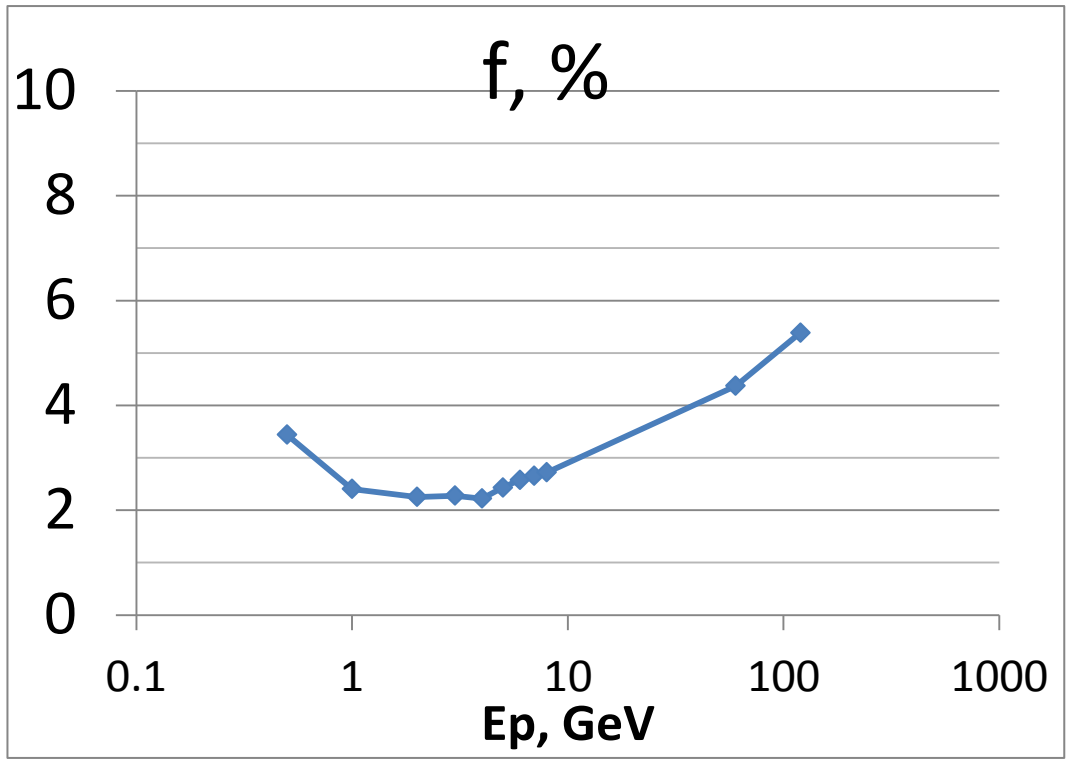

**Thermal analysis**

A thermal analysis using the ANSYS code has been carried out in order to determine the feasibility of such a target from the point of view of the heat removal. The simplest cooling scheme with the cooling lines indicated by the yellow color is shown in Figure 12. The heat map was calculated with MARS15 for a $E_p$= 3 GeV proton beam with the current $I_p$= 0.5 mA. The bunched beam was assumed to have the following parameters (similarly to those used in [11]): the bunch duration is $4 \cdot 10^{-11}$ s, the interval between bunches is $6.08 \cdot 10^{-8}$ s. The thermal analysis showed that during the first 100 s of irradiation, the target core will melt (see Figure 13). The hot spot is highly localized due to a small beam diameter as well as a low thermal conductivity of the natural uranium.

To explore a possibility to mitigate the overheating and a core meltdown issue, the beam rastering to the radius of 30 cm (instead of 5 cm) was applied to the model, and the temperature distribution in the target was studied. Figure 14 shows that the peak temperature dropped by an order of magnitude (to 3100º C) compared to the initial model. The core temperature vs time curve plotted in Figure 15 indicates that 3100º C will be reached in approximately 200 s. This temperature is still higher than the melting temperature of natU, however, it is a significant decrease relatively to R=5 cm which suggests that further beam rastering combined with scanning and adding more cooling lines can help keeping the temperature within the limits.

**Figure 12. A simple water cooling scheme. Yellow lines – water cooling lines with T=20º C.**

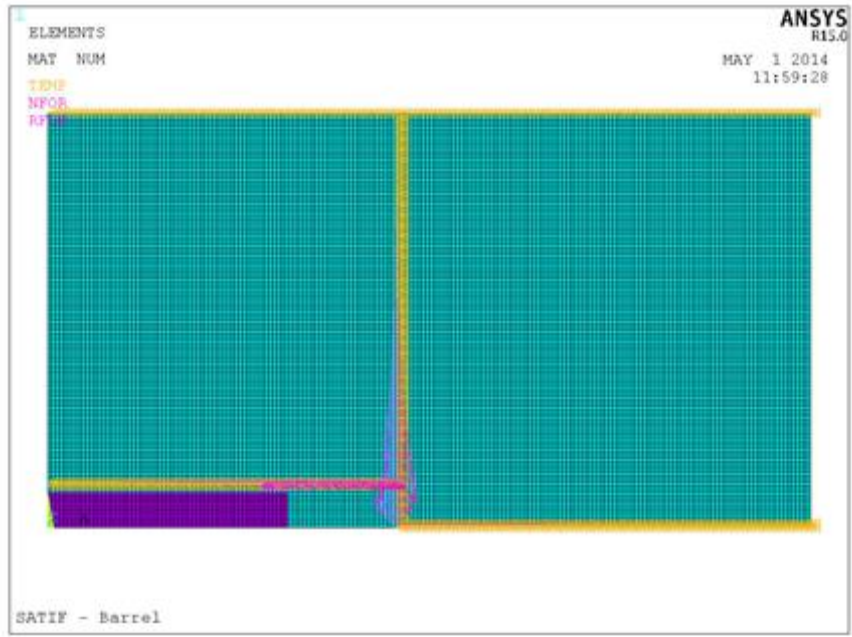

**Figure 13. Temperature distribution in the target after 100 s of irradiation with R= 5 cm beam.**

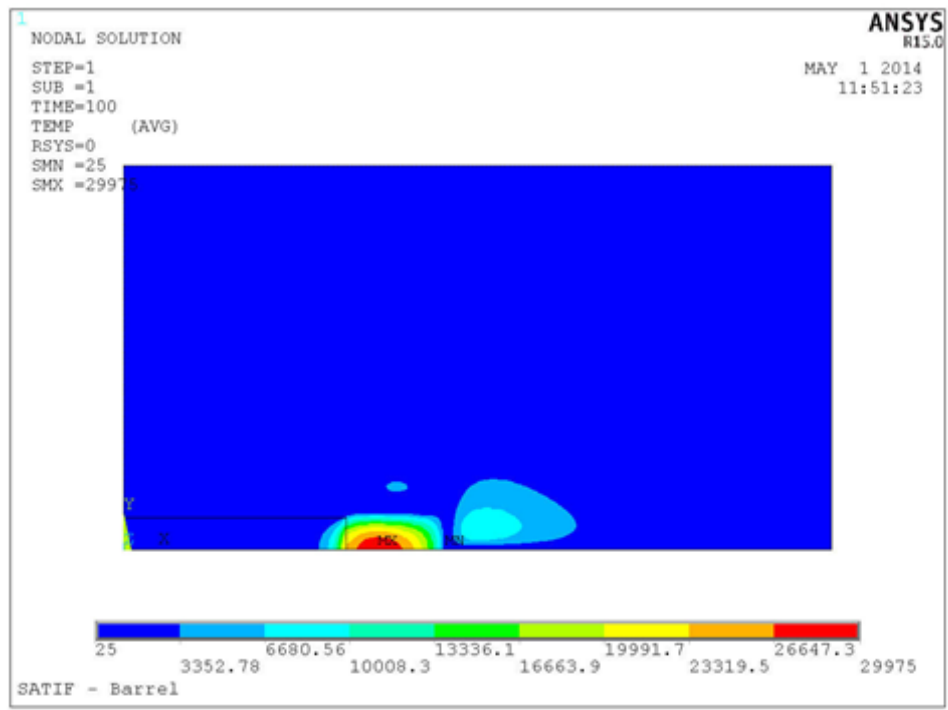

**Figure 14. Temperature distribution in the target with the beam rastered to 30 cm in radius.**

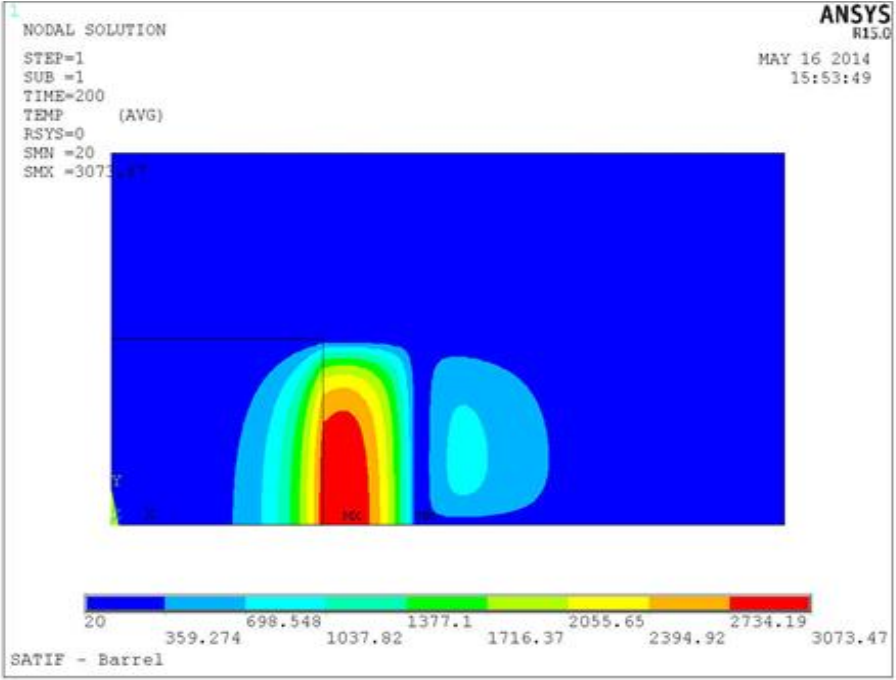

**Figure 15.  Peak core temperature as a function of time for the beam rastered to R = 30 cm.**

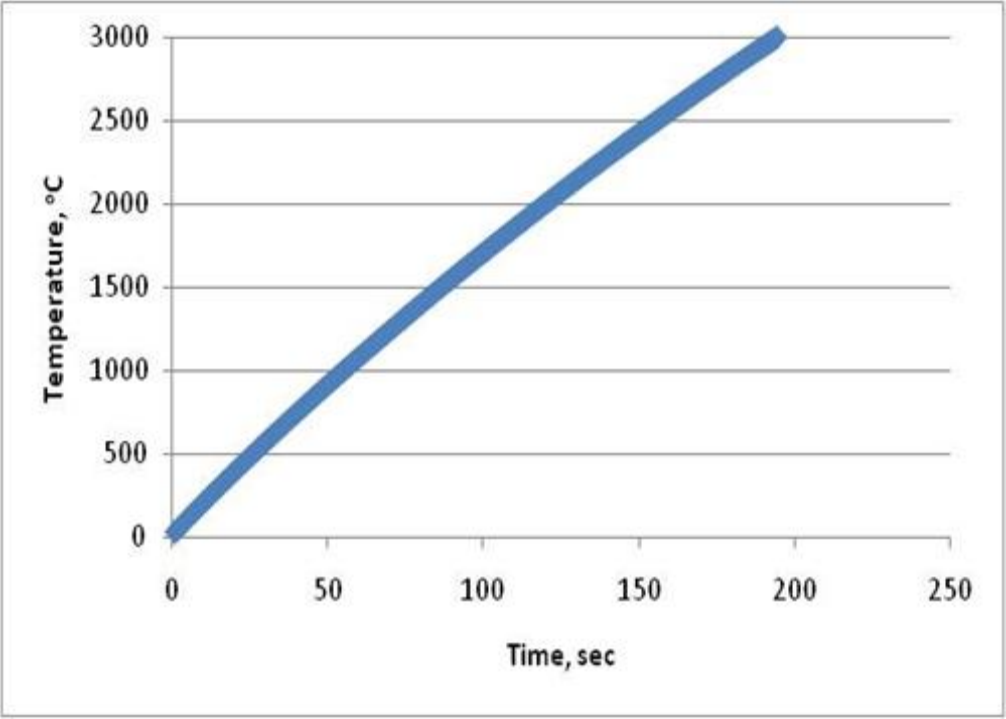

**Benchmark target**

Simulations carried out in this work suggested target dimensions sufficient to keep the neutron leakage within 5 - 7 % so that the target could perform as a full absorption one in the energy range studied. A cylindrical target with similar dimensions (R = 60 cm, L = 100 cm) was built about 20 years ago in JINR, Dubna (see Figure 16) (however, never used in experiments). Potentially, this target can be employed to benchmark neutron production predictions made in this work as well as the isotope production and fission rates. In principle, these studies could engage proton or deuteron beams of the Nuclotron accelerator of the JINR LHEP (the energy range available 0.5 - 4 AGeV, beam power ~3 W). These beams could also be used in the ADS targetry instrumentation and radiation protection research. However, for higher energy range experiments as well as for heat production studies at least an order of magnitude more beam power is required.

**Figure 16. The 21-tonne prototype $^{nat}$U target.**

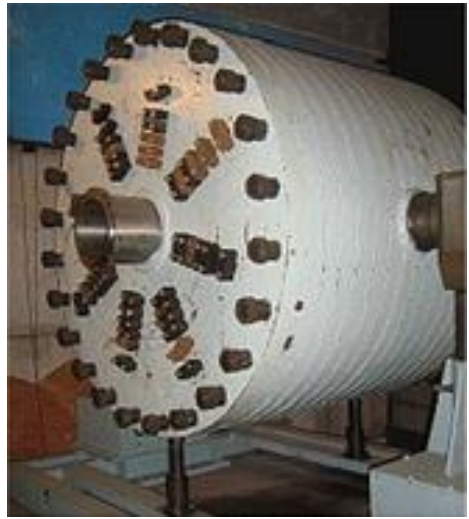

**Conclusions**

MARS15 simulations and ANSYS thermal studies of solid $^{nat}$U, Th, and W "full absorption" targets have been performed in this work. Target dimensions were optimized to keep the neutron leakage below 8% of the total number of neutrons produced in the target in the 0.5 - 120 GeV energy range. The studies reveal that in order to maximize neutron production, energy deposition, energy gain, and radiation damage the optimal energy range is 2 to 4 GeV (not 1 GeV as reported in a number of earlier works). It was shown that in the optimal energy range, the 1-mA proton beam current is sufficient to attain the 1-GW thermal output power in the case that the $^{nat}$U target is used in an ADS reactor; the fraction of the output power required to operate the accelerator in the entire energy range under scrutiny amounts to not higher than 6%.

Thermal analysis indicated that the beam on the target core would lead to a fast overheating and core meltdown. However, encouraging results were obtained by rastering the beam in a 30-cm radius (the peak temperature dropped by a factor of 10). This is a possible direction of further target optimization work. Suggested experiments with a similar existing target are able to provide data for benchmarking the simulation results discussed in this work.